\documentstyle[prl,floats,aps]{revtex}
\newcommand{\be}{\begin{equation}}
\newcommand{\ee}{\end{equation}}
\newcommand{\lll}{\langle}
\newcommand{\rrr}{\rangle}
\newcommand{\Det}{\mbox{Det}}
\newcommand{\llll}{\langle\langle}
\newcommand{\rrrr}{\rangle\rangle}
\newcommand{\T}{\mbox{Tr}\> }
\begin{document}
\hsize\textwidth\columnwidth\hsize\csname@twocolumnfalse\endcsname
\title{%
\hfill{\normalsize%
\raisebox{0pt}[0pt][0pt]%
{\vbox{\hbox{May 2001} }}%
}\\
\bf On Casimir scaling in QCD
 }
\author{\bf Vladimir Shevchenko and Yuri Simonov}
\address{Institute of Theoretical and Experimental Physics,\\
117218 B.Cheremushkinskaya, 25, Moscow, Russia.}
\maketitle

\begin{abstract}

Recent lattice calculations have confirmed that
QCD static potential for sources in different
representations of the gauge group is proportional
to eigenvalue of the
corresponding quadratic Casimir operator
with an accuracy of a few percents.
We review the present theoretical
status of the "Casimir
scaling" phenomenon
and stress its importance for analysis of
nonperturbative QCD vacuum models and other
field theories. It is argued that
Casimir scaling strongly advocates the property
of Gaussian dominance and we propose
different lattice tests to improve our
understanding of these phenomena.

\end{abstract}

\pacs{PACS numbers: 12.38.Aw, 12.38.Gc}

\section{Introduction}

There is little doubt nowadays that QCD is the true theory
of strong interactions and in perturbative domain agreement
between theory and experiment is impressive. At the same time
for nonperturbative (NP) phenomena an exact analytic formalism is still
lacking and one is dealing with a set of models or approaches,
which are not derived directly and rigorously from the QCD
 Lagrangian without any assumptions and most of these models
are introduced {\it ad hoc}.

Comparison of model predictions with experiment
allows to select models describing
one or another set of phenomena, e.g. constituent
quark model works reasonably well for hadron
spectrum, while parton model does the job for high-$Q$
events and instanton model is suitable for describing
chiral effects.
However these models are not well connected to each other
and may contradict other phenomena, e.g instanton gas
model lacks confinement.
In addition there is a class of specific models
with a special purpose to explain vacuum structure of QCD with
its property of confinement: magnetic monopoles with
abelian dominance hypothesis, center vortices model etc.

Somewhat apart from particular microscopic models a
nonperturbative approach is being developed for the last ten years
\cite{ds1} (see also review \cite{ds3}
and references therein), so
called field correlator method. It is based on
QCD and expresses all observables ultimately in terms of
gauge-invariant field strength correlators. The crucial assumption of
the method
which makes it workable is that one can restrict oneself to a few
lowest field correlators; taking only two-point Gaussian
correlator one has Gaussian stochastic model of
QCD vacuum. Extracting Gaussian correlator from lattice data
\cite{dig} one is able to predict confinement property of QCD and
calculate
quantitatively an impressive amount of data on hadron spectra and other
properties \cite{ds3,lisb,scat}.

It is clear that QCD vacuum structures derived from different
models should not necessarily be compatible with each other and
also with GSM,
while they can describe equally well some features of the
vacuum dynamics, for example actual value of string tension.
Taking gross features of the vacuum
one should distinguish two different scenarios:
one, based on an ensemble of classical solutions,
i.e. of coherent lumps of fields may be called a
coherent picture of the vacuum. Here belong
instanton model, effective abelian models, central vortex
models and some other models.
On the opposite side there is Gaussian stochastic model 
where stochastic
properties of the vacuum are taken to their extreme,
so that all higher field correlators are suppressed
and the picture of vacuum quantum correlations is close to
that of the Gaussian white noise.
It would be of great importance to understand
first of all what the real QCD vacuum is,
to which of these two different pictures it belongs.

The most important theoretical
tool to treat NP QCD available
up to now is numerical simulation of the theory on the lattice.
Confinement and other NP phenomena have been
extensively studied in this approach.
Data obtained in this way can be compared with
results of real experiments as well as with
theoretical predictions, based on particular NP models.
The latter comparison is of special interest.
Since the only  essential dynamical input
of the lattice theory is QCD Lagrangian
such comparison is important to distinguish
between models, which are in agreement
with QCD (at least in lattice  formulation) and those,
which are not.

 The property of confinement in QCD
 has two important facets.
From  observational point of view
it exhibits itself as absence of
particles, carrying nonzero color charge
as asymptotic states in Nature (in particular,
massless vectors and light spinors).
On the lattice there is another criteria
of confinement - area law:
there exists constant force (equal to $\approx$ 15 tons
in SI units) between static fundamental color charge and anticharge
at large distances (if no dynamical charges can
be created from the vacuum to screen it).
Most of NP models which can be compared
with lattice measurements
are designed to describe confinement of colour
charge and anticharge in the fundamental representation
of the gauge group $SU(3)$, i.e. the area law for the
fundamental rectangular Wilson loop and hence linearly
rising potential
between static quark and antiquark.
There is a lot of information about static fundamental
potential obtained from lattice simulations
both in quenched and unquenched approximation.
Being of great value, these data cannot answer
the questions posed above and one needs another
type of lattice experiments which were done recently with
great accuracy \cite{bali1,bali2,deldar1}.
These are the investigation of interaction
between static charges in higher $SU(3)$ representations.
In this way one can extract information about field
content of NP QCD vacuum, which is not available
if only fundamental charges are considered.

The studies in this direction have a long history (see
\cite{bali2} and references therein).
Recently
two sets of accurate lattice measurements of the static
potentials for fundamental and higher representations of the
gauge group $SU(3)$ have been
presented by G.Bali \cite{bali1,bali2} and independently by S.Deldar
\cite{deldar1}. The data strongly support the so 
called {\it Casimir scaling
hypothesis} \cite{ambjorn} which states the proportionality
of the static NP potentials for different representations
to eigenvalue of the quadratic Casimir operator
for the corresponding representation.
Actually, the level of violation of Casimir scaling (CS) 
for the string tension
is found to be at the range $5-15\%$
in the measurements \cite{deldar1} and at the level of a few
percents
in the results \cite{bali1}, while extrapolation to the continuum
limit performed in
\cite{bali2} remarkably indicated the violation of
CS law not exceeding statistical errors, of the order of
$1\%$.
All that give a hint that CS is indeed
a fundamental property of QCD
(at least on the lattice).
Surprisingly, however, the analysis of \cite{we1,we2}
has shown that to incorporate such simple
feature in some natural way is not an easy task
for most popular QCD vacuum models. Let us
make a comment about general reason for
that.
In most of coherent QCD vacuum models the corresponding
field configurations are endowed with nontrivial colour
structure, whose gauge-invariant meaning is
often not transparent.  The CS, on the other hand,
means that at least for the particularly chosen NP
quantity -- static potential -- only the simplest invariant of the
given representation, eigenvalue of the quadratic Casimir
operator, i.e. just square of the (nonabelian) charge
characterizes the potential.
Thus strongly interacting theory with complicated
vacuum paradoxically demonstrates the same scaling law of the
potential as it would be in perturbation
theory at tree level. In terms of the confining field
configuration ensemble it implies
either delicate (and as we will try to advocate in what
follows, rather unnatural)
fine tuning of the contributions coming from more
complicated color structures (higher powers of the quadratic
Casimir and higher Casimirs) or actual smallness of such structures.
The phenomenon of 
CS seen from different NP QCD models points of view
will be discussed in Section III of the present paper.

It should be said that many ideas
connected with the meaning of CS and extensively
discussed in the present paper were formulated
already in the original papers \cite{ambjorn}, \cite{belova},
\cite{bern}.
Nevertheless we find it useful to collect
them here all together with new developments
since the results from
\cite{bali2} demonstrating much higher level
of accuracy of the phenomenon are calling for reconsideration
of the theoretical background behind it.

The aim of the present paper is to draw
possible consequences from the data on Casimir scaling
constructing the most logical scheme of the
QCD vacuum structure, respecting CS.
The paper is organized in the following way:
Section II is devoted to definitions
and general discussion of the lattice data on CS,
analysis of different field theoretical
models from
CS point of view is presented in Section III,
Section IV contains our proposals of additional observables
aimed to clarify the CS
behavior and structure of the QCD vacuum,
and Section V presents conclusion and outlook.

 \section{General formalism}

The average of the Wilson loop $W(C)$
for the rectangular contour $C=R\times T$  (which we choose in
the "34" plane) for the dimension $D$ representation
of the gauge group $SU(3)$ is given by
\be
\lll W_D(C)\rrr = \left\lll {\T}_D\>{\mbox P}\exp\left(ig\int\limits_C
A_{\mu}^a T^a dz_{\mu}\right)\;\right\rrr
\label{eee1}
\ee
Here the normalized trace ${\T}_D$ is
defined as ${\T}_D {\hat1} = \frac{1}{D}  \T
{\hat 1} = 1$, fundamental generators are normalized according to
$\T T^a T^b = {\delta}^{ab} / 2$.
The $SU(3)$ representations labelled by $D=3,8,6,15a,10,27,...$
are characterized by $3^2-1=8$ hermitian generators $T^a$ which
satisfy the commutation relations $[T^a, T^b] = i f^{abc} T^c$.
 One of the main characteristics of the representation
is an eigenvalue $C_D$ of quadratic Casimir operator ${\cal C}^{(2)}_D $,
which is defined according
to
\be
{\cal C}^{(2)}_D = \delta_{ab}\> T^a T^b =
T^a T^a = C_D\cdot \hat1
\label{uu}
\ee
Our normalisation is such that $C_D = N_c$ for adjoint
representation of $SU(N_c)$, it differs by factor $N_c$ from
normalisation of \cite{hysp1}.
Since any simple algebra of rank $k$ has exactly $k$ primitive
Casimir--Racah operators \cite{cr,gelf}
of order $m_1,..,m_k$, it is possible to
express
those of higher order in terms of the primitive ones.
In the case of $SU(3)$ the primitive Casimir operators
are given by $ {\cal C}^{(2)}_D $ and
${\cal C}^{(3)}_D = {d}_{abc} T^a T^b T^c $.
The higher rank Casimir operators
are defined as
\be
{\cal C}^{(r)}_D = d^{(r)}_{(i_1 .. i_r)} T^{i_1} .. T^{i_r}
\ee
where the totally symmetric tensors $d^{(r)}_{(i_1 .. i_r)}$
on $SU(N>3)$ are expressed in terms of $\delta_{ik}$ and
$d_{ijk}$ (see, e.g.
\cite{hysp1} and references therein).
Following the notations from \cite{bali1} we
introduce Casimir ratio $d_D = C_D / C_F$, where
the fundamental Casimir $C_F = (N_c^2-1)/2N_c$ equals to
$4/3$ for $SU(3)$. Generally, for $SU(3)$
representations labelled by Dynkin coordinates
$(\mu, \nu)$ the eigenvalue of ${\cal C}^{(2)}_D$
is to be calculated according to the following
expression (see, e.g. \cite{hysp1}):
\be
C_D(\mu, \nu) = \frac{1}{3} \left(
\mu^2 + \mu \nu + \nu^2 + 3\mu + 3\nu
\right)
\label{cas2}
\ee
Notice the positivity of the quadratic Casimir
charge (it is not true, for example, for the cubic one).

The static potential between sources at the distance $R$
in the
representation $D$ is defined as:
\be
V_D(R)=-\lim_{T\to\infty}\frac{1}{T} \mbox{log}\> \lll W(C)\rrr,
\label{eq1}
\ee
We are not considering the possibility of screening
of the potential by dynamical degrees of freedom
at the moment, see discussion below.
The l.h.s. of (\ref{eq1}) can be decomposed into
two pieces of different
 physical  meaning: actual $R$-dependent potential
and self-energy part not depending on $R$,
which
explicitly contains ultraviolet
cutoff scale $a$. For the data extrapolated
to the continuum limit (see \cite{bali2}
for details)
one assumes the proper subtraction
procedure to be applied and we will
denote by $V_D(R)$  only the
former part of the potential in this case.
Notice, however that the numbers shown in Table 1
of the present paper
are based on raw lattice data from \cite{bali1} where
no any special subtraction have been performed
(a comparison shows that with the discussed accuracy the
self-energy part satisfies CS as well).

In the confinement phase lattice data are well described by
the sum of perturbative
Coulomb part,  confining linear and constant
terms:
\be
V_D(R) = \sigma_D R + v_D + \frac{\alpha_D}{R}
\label{pot}
\ee
where all coefficients are $D$-dependent.
 The  Coulomb term is now known up to two loops
in the continuum limit \cite{psr} and up to one loop on
the lattice \cite{lat} and is proportional to $C_D$ in both cases.
It is not trivial that perturbative interaction
is proportional to $C_D$ at two loops, and it is not clear
what will happen with more loops taken into account.
The CS hypothesis \cite{ambjorn} states, that
the NP confining potential is also proportional to the
first power of the quadratic Casimir $C_D$, i.e.
for the string tensions  one should have $\sigma_D/\sigma_F = d_D$.

As it was already discussed in the Introduction, this
scaling law is in very good agreement with the results
found in \cite{deldar1,bali1,bali2}. Earlier lattice calculations
of static potential between sources in adjoint representation
\cite{go1} are in general agreement
with \cite{bali1}, however, deviations from scaling
at the level of $10\%$ are
found in \cite{go2}, in particular, the value of $\sigma_8/\sigma_3$
is closer to 2 than to $9/4$ in \cite{go2} (see further
references and discussion in \cite{bali2}).

The lattice data can be understood in terms of bounds
on higher Casimir terms in the potential (\ref{pot}) \cite{we1,we2}.
Namely, one can write quite general decomposition
for the expression (\ref{pot}) of the following
form:
\be
V_D(R) = d_D V^{(2)}(R) + d_D^2 V^{(4)}(R) + ...
\label{qq}
\ee
where the dots denote terms proportional to higher powers of
$d_D$ as well as to higher Casimir ratios.
It is worth mentioning, that
since there are two independent Casimirs in $SU(3)$ case,
the series in (\ref{qq}) is in fact Taylor expansion
of a function depending on two variables (i.e. quadratic
and qubic Casimirs)
with $R$--dependent
coefficients. Notice also, that for self--adjoint
representations (those having $(p,p)$ form in Dynkin coordinates)
the cubic Casimir vanishes (see, e.g. \cite{hysp1}), so
the potential depends only on $d_D$.

The form of Taylor expansion (\ref{qq}) is strongly
motivated by perturbation theory series as well as by
cluster expansion (\ref{eq2}). However it
is not in one-to-one correspondence with
any of them. For example, the lowest quadratic
correlator $\llll FF\rrrr$ contributes only to $ V^{(2)}(R)$,
while higher ones contribute to $ V^{(2)}(R)$ as well as to higher terms.
Neither the expression (\ref{qq}) has a meaning of perturbative
series in $d_D$.

The terms $V^{(n)}(R)$ in the expansion (\ref{qq})
are representation independent. One can parametrize
the first CS violating term $V^{(4)}$ in the same way as
the total potential $V_D(R)$ as
$ V^{(4)}(R)= v^{(4)} +  \sigma^{(4)} R$,
where we omit possible CS--violating contribution to
the Coulomb potential since at small distances
its magnitude should in principle be small due to
asymptotic freedom and independent analysis
confirms this.
  Here $ v^{(4)}, \sigma^{(4)}$
   measure the $d_D^2$--contri\-bu\-tion of the
CS violating terms to the total potential (their
artificial dependence on $D$ could
come from higher powers of $d_D$, and should disappear
if all terms in the expansion (\ref{qq}) are taken
into account).
Notice, that we do not need to specify the
coordinate dependence of $ V^{(2)}(R)$.
Some results based on the data from \cite{bali1}
are presented in Table 1.
The author of \cite{bali1} used anisotropic lattice with the spatial
unit $a_s = 0.082$ Fm and anisotropy $\xi \sim 4$.
Standard $\chi^2$ fit was performed for the data in the whole
range of all measured $R$, since no fingerprint of screening is seen
up to the largest distances explored in \cite{bali1}.
Errors shown in Table 1 include statistical
and systematical ones.

\begin{table}
\centering
\caption[]{ The Casimir--scaling and Casimir--violating string
tensions
and shifts \cite{we2}.
Based on the lattice data from \cite{bali1}. All quantities
with the hats
are scaled according to $\hat u = u*10^{4}$ }
\medskip
\begin{tabular}{|ccccccc|}
\hline
$\;D\;$ & $\; {\hat\sigma}^{(4)}\;$ & $\;\Delta
{\hat \sigma}^{(4)}\;$
& $\; {\hat v}^{(4)}\;$ & $\;\Delta {\hat v}^{(4)}\;$ & $\;
\left| \sigma^{(4)}/ \sigma_D^{(2)}\right| \;$& $\;
\chi^2 / {\mbox{dof}} \;$
\\
\hline
8 &  -3.5  &   1.2 & -2.5 & 2.8 & 0.004 & 19 / 43
\\
\hline
  6 & -6.4 &  1.2 & 1.0  &  2.6 &0.007 & 26 / 42
\\
\hline
15a   &-5.2  & 0.6   &-0.6  &1.1 &0.003   &39 /
42 \\
\hline
10 &-4.9 &  0.5 & 0.2 & 1.0 &0.003 & 22 / 41 \\
\hline
\end{tabular}
\end{table}

Several comments are in order. First of all it is seen
that the CS behaviour holds with a good
accuracy, of the order of a percent
with the reasonable $\chi^2/{\mbox{dof}}$.
The ratio $\sigma^{(4)} / \sigma^{(2)}$ is less
than $1\%$ for all cases presented in Table 1.
The value of the constant term $v^{(4)}$ is
found to be compatible with zero within the error bars
for all considered $D$,
while it is not the case for $\sigma^{(4)}$.
The value of $\sigma^{(4)}$ for sextet, for example, is found
to deviate from zero at the level of approximately 
five standard deviations.
We have not found any strong systematic
dependence of
$\sigma^{(4)}$
on $D$,
which presumably confirms the validity of the expansion (\ref{qq})
and shows, that the omitted higher terms do not change the
picture in a crucial way.
The negative sign
of the $\sigma^{(4)}$ correction
is probably related in Euclidean metric
with the fact, that the fourth order
contribution is proportional to $(ig)^4 >0$ while the
Gaussian term is multiplied by $(ig)^2 <0$.

Recently the continuum limit extrapolation of the lattice data
on static potentials for different representations
was performed in \cite{bali2}.
The actual ratio
of the potentials $V_D(R) / V_F(R)$
measured in \cite{bali2}  demonstrated the deviation
from CS (i.e. from a constant equal to $d_D$)
at the level of statistical errors not exceeding
a few percents.

It is important to stress at the same time that CS property
is not universal at all possible distances.
The area law implies linear asymptotics of the potential
at large $R$. There is however an important restriction: 
even in quenched approximation with no
dynamical quarks included one should take into account
the effects of string breaking. The conventional picture
of this phenomenon suggests that in case of charges
in higher representations the parts of the potential corresponding
to the octet components of the sources are to be screened
by dynamical gluons from the vacuum and
so called gluelump states are formed eventually.
In particular, zero triality ($N_c$-ality in general case) representations
are screened completely.
 For example, in case of adjoint charges one has ${\bf 8 \otimes 8 =
27\oplus \overline{10} \oplus 10 \oplus 8 \oplus 8 \oplus 1}$ with the singlet
component. Screening for higher representations
might require more dynamical gluons.
In case of nonzero triality representations
the noncompensated  triplet is always present (for example, ${\bf
6\otimes 8 =
{\overline{3}} \oplus 6 \oplus 15 \oplus 24 }$) which makes asymptotic string
tension to be equal to the fundamental one eventually.

It is obvious that in the region where the string is broken the
expressions of the form (\ref{pot}), (\ref{qq}) have no sense.
The modification of (\ref{pot})
  due to this effect was considered in \cite{bern,ghp} in the strong coupling
expansion, resulting in the following
expression for adjoint Wilson loop in
  large $N_c$ approximation
\be
 \lll W_{adj}(C)\rrr = \exp[-{\tilde\sigma}_{adj}\cdot
{\mbox{Area}}(C)] +
\frac{\eta}{N_c^{2}}\>
\exp[- M_{gl}
\cdot {\mbox{Perimeter}}(C)]
\label{4.1}
\ee
It follows from (\ref{4.1}) that the potential (\ref{eq1})
has two different regimes at $R \le R_c$ and $R \ge R_c$
($R_c$ is the critical distance, where the second term in (\ref{4.1})
starts to dominate). Namely, one has in $T\to\infty$ limit:
$$
V(R) = \sigma R \> \theta(R_c - R) + \sigma R_c \> \theta(R - R_c)
$$
The potential becomes flat after one reaches the critical distance.
The estimates of the gluelump mass
show \cite{sim3} that the actual
value of $R_c$ is rather large, of the order of $1.5 \> {\mbox{Fm}}$ for
the $SU(3)$ gauge group in the
adjoint  case.
In other words, the lightest
gluelump is heavy in the units of $\sigma$.
 At the distances
actually explored in {\cite{bali1,bali2,deldar1} the second term
in  (\ref{4.1}) contributes at the level of lattice statistical errors.
This makes the "CS region" rather wide, at least for not
very large $D$ - from the smallest distances where tree level
perturbation theory works well to the onset of screening.

It is also worth mentioning that in the limit $T\to\infty$ the force is
discontinuous according to (\ref{4.1})
at the point $R_c$, switching from
the constant value equal to $\sigma$ at $R<R_c$ to zero value
at $R>R_c$. This {\it dynamical}
screening seems to be different from the
center vortex {\it kinematical}
screening scenario where the potential is
becoming flat with the distance smothly (see
below) even in the $T\to \infty$ limit.

Let us show how (\ref{4.1}) appears in dynamical
background perturbation theory \cite{sim3}.
We concentrate on the adjoint case
for simplicity,
generalization to other representations
can be done in the same way.
To this end we split gluon field $A_{\mu}$ as
$A_{\mu} = B_{\mu} + ga_{\mu}$, where $B_{\mu}$ represent the confining
background and $a_{\mu}$ -- the valence gluon field. Referring the reader
to \cite{sim3} for details, one gets the valence Green's function
in the background Feynman gauge in the form $G_{\mu\nu}(B) = \left( D^2(B)
\delta_{\mu\nu} + 2iF_{\mu\nu}\right)^{-1}$ and the result of integration
over valence gluons at the lowest order yields in the partition function
the factor $\left[ \Det G(B)\right]^{-\frac12}$.
The averaging in (\ref{eq1}), (\ref{eq2}) turns out to be
\be
\lll W\rrr_{B,a} = \frac{\left\lll
\left[ \Det G(B)\right]^{-\frac12}\> W(B) \right\rrr_B}{
\left\lll \left[ \Det G(B)\right]^{-\frac12}\right\rrr_B} + ...
\label{eq70}
\ee
In a similar way quark loops are accounted for by the factor
$\Det(\hat D + im)$ instead of $\left[ \Det G[B]\right]^{-\frac12}$
in (\ref{eq70}). The highest terms in $ga_{\mu}$ expansion can be
calculated systematically.

The next step is the standard loop expansion of the determinant
augmented by the world--line (Feynman--Schwinger) formalism :
\be
\left[ \Det G(B)\right]^{-\frac12}  =
\exp\left[\frac12 \T \int\limits_0^{\infty}
\frac{ds}{s} \int {\cal D}z \exp i\oint B_{\mu}dz_{\mu} \exp 2i\int F d\tau
\right] = \exp\frac12 \left\{ W_{adj}(B)
\right\}
\label{eq71}
\ee
where the curly brackets
stand for the path integral over contours, forming the loop
and the corresponding proper time
integration. Expanding the exponent in (\ref{eq71}) and keeping only the
first two terms one has
\be
\lll W \rrr = \lll W(B) \rrr_B + \frac12\> \llll
W(B) \left\{ W_{adj}(B)\right\}\rrrr_B + ...
\label{eq72}
\ee
where double brackets $\llll .. \rrrr$ denote the connected correlator.
The final step is the representation of the last term in (\ref{eq72})
as a Green's function of two gluelumps, so one
finally arrives to the asymptotic expression (\ref{4.1}).

Numerically, estimating $M_{gl}$ as 1.2 GeV for the adjoint
source \cite{sim3},
one gets for the critical distance $R_c \approx 2M_{gl}/{\sigma}_{8}
\approx 1.2$ Fm.
Therefore as it was already stressed
the preasymptotic region (where the confining potential
is linearly rising for all $D$)
is rather wide which allows to
establish CS up to the distances where signal disappears into noise on
the lattice.
Additional measurements at larger distances
could hopefully
 shed some light on the string breaking and physics of
gluelumps and establish
representation dependent bounds of the CS regions.

Let us conclude this section with the remark about
large $N_c$ limit. First of all, the "CS region"
expands up to infinity in this case, as it is clearly
seen from (\ref{4.1}). It was also advocated in \cite{ghp}
that the ratio $d_{adj}$ is equal to 2 if the limit $N_c\to\infty$
is taken (which is in line with both flux counting rule and
CS). However we do not see clear
arguments in continuum theory
why CS should become exact if $N_c \to \infty$
in general case for all higher representations.
It would be very interesting to establish in
future lattice
measurements the relative deviation from CS as a
function of $N_c$, at least for $N_c = 2..5$
(see also \cite{luc1}).
Notice, that the standard large $N_c$ factorisation
arguments are not applicable to the cumulants
entering (\ref{eq2}) since they are color-irreducible.
In particular, the proper understanding of 
Gaussian dominance property in large $N_c$ framework
has not yet been reached.

\section{NP QCD vacuum models and Casimir scaling}

As it is mentioned in the Introduction,
up to the authors knowledge there is no
example of NP QCD vacuum model
(with except of Gaussian stochastic model, see below)
which does not violate CS in "naive" sense.
All considered microscopic
models need some kind of special adjustment
in order to achieve approximate CS behavior of the static potential.
We illustrate this statement taking as examples
three popular pictures of the QCD vacuum - instanton liquid \cite{inst1},
see \cite{inst2} for review;
condensate of abelian-projected monopoles \cite{ab1}, see \cite{ab2,dig2}
for review and center vortex scenario \cite{cv1}, \cite{cv2,cv3}.
We will briefly discuss the
confining string picture, $D=3$ Georgi-Glashow model, 
MIT bag model and dimensional reduction scenario.

To begin, let us make an important remark
about the role of gauge invariance. As it was already mentioned
in the Introduction, the property of CS is essentially
nonabelian and therefore to correctly reproduce it
a NP QCD vacuum model should take into account nonabelian degrees
of freedom in a proper gauge-invariant way
(roughly speaking, the sum in expressions
like (\ref{uu})
must be taken over all $N_c^2-1$ generators).
However, it is not sufficient since even manifest
gauge-invariance of a model does not guarantee it will
demonstrate CS law. Nevertheless we find it instructive
 to separate the discussion of the models where
some particular gauge-fixing procedure plays
crucial role from those
where it does not. We will see that pattern
of CS violation is different in many respects for
these two groups of models.

\subsection{Models with the preferred choice of gauge}

As the first example of a model, based on
NP field configurations we mention here
the model of dilute instanton ensemble \cite{inst1,inst2}.
Strictly speaking this model is not expected to give
reasonable predictions for the static NP potential
since it lacks confinement and hence some essential
part of NP physics. It is instructive however to
investigate CS in this framework
since it clearly illustrates the pattern which is common for
all set of models based on ensemble of classical field
configurations.

The instanton ensemble is characterized by
averaged instanton density $n=N/V$, where
$N$ is number of instantons and antiinstantons
and $V$ - 4-volume, and mean instanton radius
$\bar\rho$.
Phenomenologically one chooses $n\approx 1\> {\mbox{Fm}}^{-4}$
and $\bar\rho \approx 0.35\> {\mbox{Fm}}$, in this case the
"diluteness parameter" or packing fraction
$n{\bar\rho}^4 / N_c $ is much less than unity,
justifying the small density expansion in the model.

The static potential in the dilute instanton gas was
calculated in \cite{gros}, see also \cite{diak2}.
In the approximation
used in \cite{gros,diak2} each instanton contribution to
Wilson loop is calculated independently, the potential
at the leading order in density is given in this case by:
\be
V(R) = \frac{1}{N_c}\; \int\limits_0^{\infty} d \rho\;
\frac{dn(\rho)}{d \rho}
\; \int d^3 z \; \T \left[ 1 - W(z,\rho,R)
\right]
\label{potent1}
\ee
where the instanton size distribution is such that
$\int_0^{\infty} dn(\rho) = n \; ; \;\; \int_0^{\infty} \rho
dn(\rho) = \bar\rho \> n$ and the
numbers of instantons and antiinstantons
are assumed to be equal: $N_{I} = N_{\bar I} = N/2$.
The potential is normalised according to the
condition $V(R=0) = 0$. In case of static
quarks belonging to higher representations eq. (\ref{potent1}) should be
generalised \cite{diak2}, with the following result:  
\be V_D(R) = 4\pi
\int\limits_0^{\infty} d\rho \frac{dn(\rho)}{d \rho}\; \rho^3
\frac{1}{d(D)}
\sum\limits_{J\in D} (2J+1) F_J(x)\;\; , \;\;\;\; x=\frac{R}{2\rho}
\label{eq20} \ee and the function $F_J(x)$ is given by some cumbersome double
integral whose exact form can be found in \cite{diak2}.  Here $d(D)\equiv D$ is
the dimension of the representation $D$ (not to be confused with
the  Casimir ratio $d_D$)
and sum over $J=0, \frac12 , 1, .. $
goes over all $SU(2)$ multiplets for decomposition of the given $SU(3)$
representation with the corresponding weights. One finds
$ d(D) \cdot C_D  = \frac{8}{3}\>
\sum\limits_{J\in D} J(J+1)(2J+1) $
and also
$\sum\limits_{J\in D}
(2J+1) = d(D)$.

 At small $x$ the functions $F_J(x) \sim
x^2$, while at large $x$ the functions $F_J(x)$ tend to $J$-dependent
constants \cite{diak2}.
 Numerically one finds at small distances
\be
V(R) = 1.79\cdot\gamma R^2 \cdot \epsilon_D + {\cal O}(R^4)
\ee
where $\gamma = \pi \bar\rho n $ 
and numerical coefficients $\epsilon_D$ for $D=3,8,10$ are given
by
$$
\epsilon_3 \; : \; \epsilon_8 \; : \; \epsilon_{10}
\;\; = \;\; 1\; : \; 1.87 \; : \; 3.11
$$
instead of Casimir scaling results
$ 1\; : \; 2.25 \; : \; 4.5 $.
Similar situation takes place
for the large distance asymptotics of the instanton--induced
potentail. It violates CS at the level of 20\% .
(see \cite{we1})

One-instanton approximation used in (\ref{potent1}), (\ref{eq20})
was recently examined on the lattice \cite{ilg1}
in order to study the instanton density dependence as well as
CS of the instanton-induced heavy quark potential.
It was found that at the distance $1.2 \>\mbox{Fm}$ the deviation
of the dilute gas formula (\ref{potent1}) from the exact
expression (\ref{eq1}) is about $10\%$ for fundamental charges
and $8\%$ for adjoint charges at the density $n=0.2\> {\mbox{Fm}}^{-4}$,
while it increases up to $\sim 20\%$ for fundamental charges
and $\sim 50\%$ for adjoint charges at the
phenomenological density $n=1\> {\mbox{Fm}}^{-4}$.
Moreover, linear density dependence of the potential
extracted from the Wilson loop (\ref{potent1})
breaks down for the densities $\sim 0.5\> \mbox{Fm}^{-4}$
at the distance $1\> \mbox{Fm}$. The
ratio of the potentials obtained according to
(\ref{potent1}) is found in \cite{ilg1}
at different distances and densities (to be compared
with the Casimir ratio $d_8 = C_{adj}/C_F = 9/4 = 2.25$)
and it
goes as $2.12$ at $R=0.4\>\mbox{Fm}, \; n=0.2 \> {\mbox{Fm}}^{-4}$;
$1.91$ at $R=0.8\>\mbox{Fm}, \; n=0.6 \> {\mbox{Fm}}^{-4}$;
$1.49$ at $R=1.2\>\mbox{Fm}, \; n=1.0 \> {\mbox{Fm}}^{-4}$.
The ratio of the potentials
extracted from (\ref{eq1}) has been found in \cite{ilg1}
to be equal to 2 with an accuracy of a
few percents for all explored distances.

All these results make it clear that
instanton ensemble cannot be a natural
framework for explanation of CS.
It also demonstrates that if
one is considering field configurations on the lattice
which by some reasons
are mostly made of instantons (as it happens, for example,
after the cooling procedure),
CS does not hold in such vacuum.
It would be interesting to observe directly on the
lattice the CS breaking in this case.

Let us mention another interesting problem
in this respect -- the pattern of CS in SUSY QCD.
One should expect the role of instantons in vacuum
dynamics of ${\cal N}=2$ and even ${\cal N}=1$
supersymmetric Yang-Mills theory
to be much more transparent than in real world
QCD.  On general grounds one expects the simpler structure of perturbation
theory and hence the absence of higher Casimirs in
the perturbatively induced potential. On the other
hand, coherent NP vacuum made mostly of instantons and antiinstantons
should produce significant violations of CS for the NP part of the static
potential.

Another aspect is related to elementary
$k$-strings, which can be formed between static sources
for the gauge groups $SU(N_c)$ with $N_c>3$.
Such strings are stable and characterized by the
string tension $\sigma_k$ (see, e.g. \cite{stras1}).
Recent lattice studies
in this direction \cite{luc1,luc2,oh}
are of much interest. They are aimed to test CS
in a different setup from that of \cite{bali1,deldar1}
and compare the results
with alternative predictions, e.g. with the so called
MQCD conjecture \cite{stras}, see also \cite{stras1}.
This project is far from being complete and
one needs more accurate data to make reliable conclusions,
In particular, MQCD predicts $\sigma_k \propto \sin(\pi k /N_c)$,
which gives for $\sigma_2 / \sigma_{fund}$ the ratios
$1\; : \; 1.414.. \; : \; 1.618...$ for $N_c = 3,4,5$,
respectively. On the other hand CS hypothesis predicts
that this ratio should scale as $1\; : \; 4/3 \; : \; 3/2$.
It would be very interesting to test these predictions
in further lattice calculations (see \cite{luc2}
and references therein).

Coming back to the real QCD vacuum, one can
guess two possible explanations of the fact that
instantons do not destroy CS.
Either instantons are strongly suppressed in the real (hot)
QCD vacuum (as it was observed in different respect
in \cite{pv})
while they are recovered by the cooling procedure.
Or else instanton
medium is dense
 and strongly differs from dilute instanton gas, in such a
 way that higher cumulant components of such collectivized instantons
are suppressed. Interesting to note, that linear confinement missing
in the dilute gas, would be recovered in this case.

Now we come to another NP QCD vacuum model,
based on the center dominance idea. An
important role of the center of the gauge group
was stressed by 'tHooft in \cite{cv1}.
A center vortices model tries to understand confinement
in terms of interaction of the current forming
the Wilson loop with the topologically
nontrivial field configurations charged with respect to the
center of the gauge group.
A center vortex is a topologically stable field configuration
with the topology of the surface, which carries (chromo)magnetic flux and
interacts with the external (chromo)electric current
in the following way:
\be
W(C) \to \exp\left(2\pi i n L(C,S) / N\right) \; W(C)
\label{cv1}
\ee
where $n=1,..,N-1$ and $L(C,S)$ represents the linking
number (in four dimensions) between Wilson contour
$C$ and vortex surface $S$.
The idea of center dominance implies that dominant contribution
to the string tension comes from the quantum fluctuations
in the number of center vortices linking the loop.
The model predicts the following static potential
(the reader interested in the details of the center
vortex model is referred to \cite{cv2,cv3} and references therein):
\be
V_D(R) = - \sum\limits_{n=1}^{N-1}\;
\mbox{log}\> \left(1 -  f_n[1- \mbox{Re}\>
G_D({\vec\alpha}^n_C(x))]\right)
\label{cv2}
\ee
where the function $ G_D({\vec\alpha}^n_C(x))$
is defined as $G_D({\vec\alpha}^n_C(x)) = (1/d(D))\>
\T \exp[i\vec\alpha {\vec H}]$ and $\{ H_i \}$
is the set of generators from the Cartan subalgebra.
The function ${\vec\alpha}^n_C(x)$  represents
the corresponding  solid angle
and depends both on the Wilson contour $C$ and on the
position of the vortex center $x$. The parameter $f_n$
determines the probability for the vortex to cross with
the contour.
The expression (\ref{cv2}) is derived in the so called fat
center vortex model, where each vortex has finite thickness.
This parameter is crucial for potential to be nonzero
at intermediate distances in the case of
zero triality representations: in the limit of vanishing thickness
adjoint (and all zero $N_c$-ality representations)
loop has no interactions with center vortices.
So the introduction of thickness leaves a window for CS,
which is essentially perimeter-type effect in this model
\cite{cv3}.
The price to pay, however, is the
contradiction between (\ref{cv2}) and (\ref{4.1}), in
particular, the potential in the adjoint representation
is not linear (and even not a convex \cite{deldar3})
function of the distance, despite
at some interval of distances it
can be approximated by linear term
and its flattening
is going smoothly with the distance in this model, to be
compared with (\ref{4.1}).

The exact value of the
CS violation is strongly model dependent in this approach which
makes it difficult to put stringent bounds on the parameters of fat vortices
model based on CS.
For example, in the $SU(2)$ case  the deviation
of the center vortices induced potential from CS behaviour is about $30\%$ for
$j=1$ (adjoint)
and $\sim 80\%$ for $j=3/2$ at large distances \cite{cv3}, but presumably
it can be made much smaller by the proper adjustment of parameters.

Lattice measurements of the potential for different
representations of $SU(3)$ were performed in center vortex
model in \cite{deldar3}.
In is instructive to consider normalised Casimir ratio
\be
\xi_D (R) = \frac{1}{d_D} \; \frac{V_D(R)}{V_F(R)}
\label{oo}
\ee
In the theory with exact CS one expects $\xi(R) \equiv 1$.
If we take for $R=R_0$ some fixed point,
for example such that $R_0 V_F(R_0) = 2.5$ (this
point corresponds to the edge of the region,
where quasilinear asymptotics of $V_F(R)$ begins),
we get from the results \cite{deldar3} for some of the
representations:
$$
\xi_8(R_0) = 0.98 \>, \;
\xi_{27}(R_0) = 0.83 \>, \;
\xi_8(2 R_0) = 0.86 \>, \;
\xi_{15a}(2 R_0) = 0.63 \>, \;
\xi_{10}(3 R_0) = 0.47
 $$
These numbers clearly demonstrate the level of the CS
violation in the fat center vortex model. It was suggested
in \cite{cv3} that the model can be tuned
in order to explain properly the CS at intermediate
distances, in particular, by adjusting the vortex
profile function. At the same time, the initial
root for CS violation lies in this case in
the lack of the gauge-invariant formulation of the
model. It would be
interesting to understand in this respect
the gauge-invariant contents if any of the notion
of center dominance.

It is of interest to discuss the situation
which takes place in another popular
confinement scenario - dual Meissner effect
in the abelian-projected formulation of QCD \cite{ab1,mand3}.
We refer the reader to the review papers \cite{dig2,ab2},
where the details of this approach are extensively
discussed. It is worth noting that at the present
moment the dual Meissner confinement scenario
is often considered as the most reliable microscopic
picture of confinement and the method
of abelian projections (in line with the abelian
dominance hypothesis \cite{abdom}) as the most adequate
language for its description.
As a warm-up example, let us consider first $D=3$ Georgi-Glashow
model. The $SU(2)$ model contains nonabelian vector field
with two components $A_{\mu}^{\pm}$ acquiring nonzero mass in
the Higgs phase, while the third component $A_{\mu}^3$
remains massless and provides dominant contribution
to the Wilson loop. Existence of monopoles
in this model leads to area law \cite{pppp}.
The CS property of Georgi-Glashow theory in comparison with
abelian-projected QCD was discussed in \cite{deldeb}.
In the former case one has
$$
\lll W_j(C)\rrr =  \left\lll \T_j\>{\mbox P}\exp\left(ig\int\limits_C
A_{\mu}^i T^i dz_{\mu}\right)\;\right\rrr  \sim
\left\lll \T_j\> \exp\left(ig\int\limits_C
A_{\mu}^3 T^3 dz_{\mu}\right)\;\right\rrr  =
$$
\be
= \frac{1}{2j+1}\>\sum_{m=-j}^{j}\> \left\lll \exp\left(igm\int\limits_C
A_{\mu}^3 dz_{\mu}\right)\;\right\rrr
\label{gg}
\ee
The average for large contours is dominated by $m=0$ or $m=\pm
\frac12$ for $2j$ even and odd, respectively. It is clear therefore
that e.g. zero charge part of diagonal component of 
adjoint external current remains
unconfined and hence the model is CS violating.

In terms of gauge-invariant correlators given by
$\lll H_{\mu_1}(k_1) .. H_{\mu_n}(k_n) \rrr$ where 
one defines abelian field strength $H_{\mu} =
{\epsilon}_{\mu\nu\rho}{\phi^a}{F^a}_{\nu\rho}/m_W
$, where $\phi^a$
is
adjoint scalar Higgs field and $m_W$ - the mass of
$A_{\mu}^{\pm}$ components of full nonabelian gauge field,
the vacuum of Georgi-Glashow model
made of 't Hooft-Polyakov monopole-antimonopole pairs is coherent
and not stochastic.
Indeed, nonperturbatively generated mass scale $M \sim
\exp(-\epsilon m_W/2 e^2)$ characterizing
exponential falloff of the correlators in $x$-space
enters string tension in
highly nontrivial nonlinear way,
while extracting confining part of the two-point
correlator $\lll H_{\mu}(k)H_{\nu}(-k) \rrr$ one gets
simply
$D(k)= M^2/(k^2 + M^2)$ \cite{dima2}. It is worth saying that
nonperturbative physics of $D=3$ Georgi Glashow model
is investigated on the lattice in details (see, e.g. \cite{deldeb}
and references therein) and conclusions of theoretical analysis
performed in \cite{pppp} have been supported.

We have learned from this example that Higgs mechanism
may be a source of CS breaking since it makes
some gauge degrees of freedom massive
without destroying of gauge invariance,
while other ones remain massless. This pattern is
supposed to occur in QCD at finite density and we
will briefly discuss it at the end of Section IV.
Now we are coming back to abelian-projected QCD.
Application of the method
of abelian projections crucially depends on
adopted gauge fixing procedure, so one could expect problems
with the explanation of CS in this framework.
Indeed it happens to be the case \cite{smi}.
Abelian dominance hypothesis \cite{abdom} in QCD
states, roughly speaking, infrared dominance of diagonal
components $A_{\mu}^3 , A_{\mu}^8$ of the
full nonabelian gauge field $A_{\mu}^a$ and hence
suggests to omit nondiagonal parts in (\ref{eee1}) in QCD
just as it is done in (\ref{gg}).
Doing so, one immediately obtaines deconfinement
of adjoint charges at all distances in plain contradiction
with the lattice data \cite{deldeb}.
It indicates that the dynamics
of long-range QCD is not like that of duplicate
compact QED (in contrast with Georgy-Glashow model,
which behaves as compact QED at the considered limit) and
color gauge symmetry is not broken by some composite Higgs
field (as it effectively happens in abelian dominance
scenario in QCD).

There exist a few approaches in the literature aimed to
cure this problem in the framework of abelian projection.
All of them try to take into account the nonabelian
degrees of freedom in more or less sophisticated manner.
In particular, the reader is referred to
\cite{poulis} where the $m=0$ component of
adjoint source is effectively dressed with a virtual
cloud of charged degrees of freedom and hence
is able to interact with neutral "photon".
Adjoint string tension arises in this picture
at intermediate distances due to interaction
of diagonal abelian projected gluons with the part of the
adjoint source doubly charged with respect to the Cartan
subgroup. If one naively omits the corresponding Faddeev--Popov
determinant it gives $\sigma_{adj} = 4 \sigma_{fund}$, i.e.
just the square of the abelian double charge.
 It is expected that loop expansion of the determinant
produces terms, correcting the above behaviour to the Casimir
scaling ratio, but it has never been shown
explicitly.

Another approach was discussed in \cite{toki}.
The results of \cite{toki}
are based on the so called Weyl symmetric
formulation of
effective dual abelian Higgs
model (DAHM) with the direct summation
over $SU(3)$ root vectors.
The following prescription for the external
current interaction term was adopted (see also \cite{abdom})
\be
\bar\psi \gamma_{\mu} T^a A_{\mu}^a \psi
\to
\bar\psi \gamma_{\mu} (T^3 A_{\mu}^3 +
T^8 A_{\mu}^8) \psi
\label{cuo}
\ee
with subsequent recipe $ \bar\psi(x) \gamma_{\mu} T^{3,8} \psi(x)
\to Q^{3,8}\cdot\oint_C \delta(z-x) dz_{\mu}$.
The above ansatz is different from (\ref{gg}) since it
implicitely assumes some kind of averaging over different
color components of external current and consequently
produces (contrary to (\ref{gg}))
area law for Wilson loop in all representations.
It was found in this framework in \cite{toki}
that CS favours different values of
Ginzburg-Landau parameter for different representations,
lying in the Type II range $\kappa = m_B/m_{H} = 5 \sim 9$.
For $\kappa = 1$
flux tubes become noninteracting and one has
flux counting rule $\sigma_D / \sigma_{fund} =
\mu+\nu$, where $\mu , \nu$ are Dynkin weights
of the $SU(3)$ representation (see
\cite{toki} and references therein).

The formalism proposed in \cite{pk1} deals with 
some lattice motivated reformulation of QCD,
where a special kind of averaging over  
different
choices of ${[U(1)]}^2$ subgroups of the original
$SU(3)$ is implemented
instead of having one fixed choice.
The situation with CS in this model if realistic 
values of Ginzburg-Landau parameter $\kappa$ are taken
(e.g. from fitting of QCD string profile by 
DAHM induced formulas \cite{pk4}), is still unclear.
We refer the reader to
the review \cite{ab2} where this set of problems is discussed.

Concluding the discussion of abelian projected theories
let us mention that it could be of interest to calculate 
the potential in approaches advocated in  \cite{poulis,toki,pk1}
at all distances and for different representations
of $SU(N)$
and directly compare it with the results
of \cite{bali2,deldar1}. As it is clear from the above 
discussion the problem to reproduce CS is still
far from being closed for abelian projected theories.

All the examples discussed above
clearly demonstrates
crucial importance to count properly all relevant
nonabelian degrees of freedom.
Any NP QCD vacuum picture analysed so far
lacking manifest gauge-invariance had encountered
internal difficulties explaining CS.
In the next section we will consider
another class of models where the
nonabelian gauge invariance is kept intact from the
beginning.

\subsection{Gauge-invariant models}

We start with MIT bag model \cite{mit}
and refer the reader to papers \cite{thorn}
where the problem of the CS in this model was
addressed. It was advocated the following
relation between string tensions in the MIT
bag model: $\sigma_D/\sigma_F = \sqrt{d_D}$.
It is easy to see, that this relation
is in contradiction with the data from \cite{bali2}
at the level of a few dozen standard deviations.
Qualitatively, it comes from the fact, that bag
model misses the string, a crucial ingredient of
modern picture of NP QCD, based on
lattice results.
The string in the static charge--anticharge case
has two ends, with the factor $gT^a_D$ attached to each one,
and fixed, representation--independent radius (see below),
so the total contribution to the energy in proportional to the
charge squared, i.e. $C_D$. At the same time,
in the MIT bag model the charge enters the energy
linearly.

Another family of models to be considered here are the
so called "confining string" models. Attempts to build a
theory of the
confining string have begun more than 30 years ago (see e.g.\cite{polyakov}).
Till the present moment however there is no selfconsistent
quantum string theory applicable to QCD string, those formation
in the confinement phase is clearly seen on the lattice.
Moreover, such theory is absent even for abelian Higgs model,
where the strings exist as classical solutions
(Abrikosov-Nielsen-Olesen strings) (see \cite{dima1} and references
therein).
As an example let us consider here the static potential
induced by quantum Nambu-Goto string.

The action of the model
is proportional to area of a surface
bounded by the static sources worldlines. Quantum dynamics of this surface
produces additional contribution to the confining
potential besides the leading linear term \cite{lush}:
\be
\sigma R \to \sigma R - \frac{\pi}{12} \> \frac{1}{R} + ...
\label{eq7}
\ee
where the term $-\pi/(12\cdot R)$ will be
referred to as the String Vibration (SV) term.
Despite the Nambu--Goto string
model cannot be rigorously defined in $D=4$ (see e.g.
\cite{polyakov}), and, in particular
the expansion of the r.h.s. of (\ref{eq7})
is valid  at large distances $R$ only,
it is instructive to look whether or not the lattice data
support the existence of such term. It is also worth noting, that
the dimensionless coefficient $-\pi/12$ is universal and
determined by the only two factors: target space dimension
and the chosen string model. Having both factors fixed, it
cannot be freely adjusted.
Assuming $\sigma_D = d_D \sigma_F$, it
is easy to see, that the
Nambu--Goto SV term violates CS of the potential,
as it was noticed already in \cite{ambjorn}.

It is a nontrivial task
to separate the contributions of the discussed sort in the
confining potential as it is because these corrections
are essentially large distance effect, where they are
subleading.
But they have to become pronounced in the expression (\ref{qq})
due to scaling violation.
Namely, assuming that SV term is the only one violating CS,
one obtains
\be
{\frac{V_D(R)}{V_F(R)} - d_D } =
\frac{d_D - 1}{V_F(R)R_0}\> \frac{\pi}{12}
\>\frac{R_0}{R} + ..
\label{eq45}
\ee
where the dots denote terms, omitted in (\ref{eq7})
and $R_0$ is some arbitrary scale, for example Sommer scale.
On general grounds one should expect, that string picture works at distances
$R \gg 1/\sqrt{\sigma_D}$
since no other dimensionful parameter enters the Nambu-Goto string action
(in particular, string in this model is infinitely thin).
We expect therefore, that the data \cite{bali1,bali2} allow to extract the
possible contribution
from CS-violating SV term at intermediate distances $\sim 1$ Fm.
The l.h.s. of (\ref{eq45}) is compatible with zero at the level of $1\sigma$
from the data \cite{bali2} while the r.h.s. of (\ref{eq45})
is nonzero and is rising with $d_D$. However the
statistical errors of the expression on the l.h.s.
at such large distances are too big to
make definite conclusion about ruling out
the SV term of the discussed form. Additional accurate
measurements are needed for this purpose.
Notice, that the
sign of the Nambu--Goto SV correction is opposite
to what was found for CS violation in
\cite{bali1}.

One way to combine CS and string dynamics was proposed
already in \cite{ambjorn} -- just to multiply
all the potential (\ref{eq7}) with $d_D$,
accounting for "$C_D$ elementary fluxes"
attached to the static sources.
It should be stressed that it implies different
physical mechanisms responsible for
creation of the string and its quantum fluctuations
and presents actually a model, different from Nambu-Goto
string.
The situation with direct lattice measurements of SV corrections
is not yet clear. While the authors of \cite{green}
claim the disagreement between hybrid spectrum and the string picture,
but there are also evidences in favor of SV term \cite{go4}.
The question certainly deserves futher study.

From theoretical point of view it stays to be proved
that the QCD confining string and the
simplest bosonic Nambu--Goto string model belong to the same
universality class, see e.g. recent discussion in \cite{dubin},
and the theoretical
background of (\ref{eq7}) is not yet clear.
There are also arguments against this possibility
(see, e.g. \cite{polyakov}).
The theory of the QCD confining string -- whatever it will be --
must explain the observed Casimir scaling of the potential
at intermediate distances.
There are arguments \cite{dubin}
that the confining string in the strongly coupled
continuous Yang-Mills
theory posessess two limiting regimes: CS
respecting, which occurs if the area of the
surface bound by the Wilson loop
is smaller than some critical area, and flux-counting
regime, which starts to dominate for
asymptotically large surfaces.
It could happen that strings created by the
sources in higher representations would break at
smaller distances than those where the
quantum fluctuations of the string become visible.
In this case CS region will cover all the scales
where the string exists as a stable object
and no room to detect any significant
contribution from SV terms will be left.
This problem is absent for stable $k$-strings
on $SU(N_c>3)$ (see above), and it is reasonable to
expect CS violation at large distances in this case
caused just by SV corrections, i.e.
by quantum dynamics of the string.

Let us mention two regimes of the $SU(N)$ gauge theory where
the CS property is under theoretical control. The first
one is two-dimensional Yang-Mills theory. In $2d$ case
the theory has no propagating degrees of freedom and the
static Coulomb potential which is linear in two
dimensions is exactly proportional to $C_D$. Another
example is the strong coupling regime, where the partition function
of the $4d$
lattice Yang-Mills theory is dominated by the
two--dimensional surfaces with the bare string tension proportional
to the quadratic Casimir.
It is worth reminding that
the original motivation
for invention of the CS hypothesis \cite{ambjorn} was based on
the dimensional reduction scenario \cite{belova},
which assumes that
there exist nontrivial relation between NP
observables, e.g. string tensions
in $2d$ and $4d$ Yang-Mills theories, if
proper identification of the couplings is done.
QCD in $2d$ and $4d$ have indeed
much in common while also some important differences.
One of them to be mentioned here is the Lorentz
structure of NP linear confining potential
which is vector in $2d$ but scalar in $4d$.
We will briefly discuss the relation of this picture
with Gaussian dominance in Section V.

In the gauge--invariant NP background field
formalism (see review\cite{ds3})
the CS property
has two possible interpretations.
To understand the corresponding physical pictures,
let us formulate some basic ideas of field correlator method.
The approach decomposes the gluon field into perturbative
field and nonperturbative background with the latter to
be taken into account by means of gauge-invariant
irreducible correlators (cumulants)
$\lll \lll \T \Phi(x_0,x_1)F(x_1)\Phi(x_1,x_0) ...
\Phi(x_0,x_n)F(x_n)\Phi(x_n,x_0) \rrr \rrr$.
In particular, the Wilson loop average in the $"34"$ plane
 admits following expansion (see \cite{ds3} for details)
$$
\lll W_D(C)\rrr = \left\lll {\T}_D\>{\mbox P}\exp\left(i\int\limits_C
A_{\mu}^a T^a dz_{\mu}\right)\;\right\rrr =
 \left\lll {\T}_D\>{\cal P}\exp\left(i\int\limits_S
F^a(u) T^a d\sigma(u) \right)\;\right\rrr
=
$$
\be
= \T_D {\cal P}_x\>\exp \>\sum\limits_{n=2}^{\infty}\int\limits_S i^n \>\llll
F(u^{(1)})
.. F(u^{(n)}) \rrrr d\sigma(u^{(1)})... d\sigma(u^{(n)})
= \T_D \exp \>\sum\limits_{n=2}^{\infty}
i^n  \>{\Delta}_D^{(n)}[S]
\label{eq2}
\ee
The nonabelian Stokes theorem
\cite{nast} and the cluster expansion
theorem have been used in deriving (\ref{eq2}).
Here
$F(u)d\sigma(u) =
\Phi(x_0,u)E_3^a(u)T^a\Phi(u,x_0)
d \sigma_{34}(u)$,
where $\Phi$ is a phase factor,
\be
\Phi(x_0, u^{(k)}) = \mbox{P}\exp \left[ i \int\limits_{x_0}^{u^{(k)}}
A_{\mu}(z) dz_{\mu} \right]
\label{transp}
\ee
$u^{(k)}$ and $x_0$ are the points on the surface $S$
bound by the contour
$C=\partial S$.
The double brackets $\llll ... \rrrr$
denote irreducible Green's functions
proportional to the unit matrix in the colour
space (and therefore only spacial ordering
${\cal P}_x$ enters (\ref{eq2})).
Since (\ref{eq2}) is gauge-invariant, it is convenient
to make use of generalized contour gauge \cite{gcg}, which
is defined by the condition $\Phi(x_0, u^{(k)}) \equiv 1$.

In the confinement phase vacuum is disordered
in a sense that averages from the r.h.s. of (\ref{eq2})
develop finite correlation length
$T_g$, which for the lowest
two-point correlator was found on the lattice to be rather
small \cite{dig}: $T_g \approx 0.2$ Fm in the quenched approximation
for $SU(3)$ (to be compared with the old
stochastic proposal of \cite{olesen}).
The Gaussian dominance  \cite{ds1}
implies that the dominant contribution to the potential
(and most of other observables)
comes from the lowest two-point correlator $\lll FF \rrr$
if the integration surface in (\ref{eq2}) is taken as minimal.
It is easy to see that
\be
\lll \T_D \> F(1) F(2)\rrr =
\frac{C_D}{N_c^2 -1}\> \lll F^a(1)
F^a(2)\rrr
=
\frac{d_D}{2N_c}\> \lll F^a(1)
F^a(2)\rrr,
\label{eq4}
\ee
so Gaussian approximation provides exact CS.
Notice that (\ref{eq4}) holds true since
parallel transported field strength tensors 
$F(u)= \Phi(x_0,u)F^a(u) T^a \Phi(u,x_0)$
from (\ref{eq2}) obey the same standard
commutation relations as
bare $F_{\mu\nu}^a(u) T^a$ do.

The CS property of Gaussian vacuum depends neither
on the actual coordinate profile of the potential
nor on that of the correlator, in particular,
it holds for any value of $T_g$.
It is true also for representations where
the linear potential is just some kind of
intermediate distance behavior and changes profile
at larger $R$.
The coordinate dependence
of the potential,
not directly related to CS,
can be analysed at the distances small enough
not to be affected by screening effects.

The average $\lll W_D(C) \rrr$ does not depend
on the choice of the surface $S$ in (\ref{eq2}).
At the same time the particular choice of $S$
determines the weight of integral contribution
of each cumulant
${\Delta}_D^{(n)}[S]$
to the Wilson loop average,
and this weight is obviously $S$-dependent.
However this dependence is cancelled out
if the whole sum in the r.h.s. of (\ref{eq2})
is taken \cite{nast}. It is clear therefore that for
surfaces strongly different from the minimal
one there is some kind of cancellations
between higher terms, while for the minimal
surface CS might be either
a result of cancellations between higher
Casimir contributions from higher
cumulants or just follow from the Gaussian dominance
(terms with $n>2$ in (\ref{eq2}) give small contribution).
We will discuss below possible ways to check that.
In either case, we should stress again
that Gaussian dominance is understood
throughout this paper in the following sense:
if $S=S_{min}$ is the minimal surface,
${\Delta}_D^{(2)}[S_{min}]$ in (\ref{eq2})
is considerably larger than all other terms.
The gauge-invariance of
${\Delta}_D^{(n)}[S]$ makes the notion of Gaussian
dominance gauge-invariant as well.

\section{What can we learn from the CS?}

In this section we discuss other
consequences of CS and try to argue that
some further detalisation of the CS property
in lattice calculations might be of much use
for our understanding of the strong interaction physics
and the mechanism of confinement.

\subsection{Three-point correlators}

Only two-body potentials
of the meson type
have been discussed till the present moment.
It is also of interest to study
three-body system, which in fundamental and
adjoint cases corresponds to baryons
and tree-gluon glueballs, respectively.
The tree-body potential for fundamental sources
$V^{3Q}(R)$ is defined
by expression (\ref{eq1}) where $W^{(3Q)}$ is taken to be
\be
W^{(3Q)} = \epsilon_{\alpha\beta\gamma}
\Phi_{\rho}^{\alpha}(C_1)
\Phi_{\sigma}^{\beta}(C_2)
\Phi_{\eta}^{\gamma}(C_3)
\epsilon^{\rho \sigma \eta}
\label{bar}
\ee
and the contours $C_1 , C_2 , C_3 $ are formed by
static quark trajectories.
Indices of the antisymmetric tensors $\epsilon_{\alpha\beta\gamma} $
run from 1 to 3. For ajoint sources the
corresponding expression is given by
\be
W^{(3G)} = \Omega_{abc}\>
\Phi^{ad}(C_1)
\Phi^{be}(C_2)
\Phi^{cg}(C_3)
 \Omega_{deg}
\label{bar1}
\ee
where the tensors $\Omega_{abc}$ determine the
colour structure of the $3G$ state corresponding to the
minimal energy and adjoint indices run from 1 to 8.
Let us take spacial configuration of the system
 to be equilaterial triangle with the side $R$.
Naively one might assume the same CS law for the
3-body case, i.e. $ V^{3G}(R) /
V^{3Q}(R) = C_{adj} / C_F $.

However, one has to conclude that
the CS hypothesis in 3-body case should fail.
The reason lies in different geometry of the 2-body and
3-body systems. Indeed, in case of two static sources
the minimal area or the surface bound by the Wilson contour is
provided by the plane, one and the same for all representations.
In 3-body case in adjoint (and higher) representation
the string may go either along Mercedes-star
configuration (with the formation of a
string junction, Fig.1a) or as triangle
configuration (see Fig.1b) and one can easily argue that the energy of
the latter variant is lower (notice at the same time that
there is no such choice in the case of fundamental sources).
Indeed, for the static (i.e. noninteracting and nonvibrating)
strings connecting adjoint sources
one has $V^{3G}_{\triangle}(R) =
\sigma_{F} \cdot 3\>R$ for the triangle configuration
and $V^{3G}_{\bot}(R) = \sigma_{adj}\cdot 3\cdot(R/\sqrt{3}) =
(9/4)\sigma_{F} \cdot \sqrt{3}\> R > V^{3G}_{\triangle}(R)$
for the Mercedes-star case, where CS relation
$\sigma_{adj} = (9/4) \sigma_{F}$ was used.
For higher representations one can imagine more complicated
geometry of strings, for example
with more than one junction.
Therefore we do not expect that lattice analysis of
the 3-body potentials for higher representations
will indicate CS in the same fashion as it does for
2-body potentials. In particular, it is interesting to check
the ratio
$V^{3G}(R) / V^{3Q}(R) \simeq \sqrt{3}$ which would indicate
different geometry of the strings (and at the same time
indirectly support CS law) in the two cases.

\subsection{Stochastic vs coherent}

It was shown in the previous section how
the gauge-invariant picture of Gaussian
dominance explains CS of static potential.
It was also stressed that the CS property on the other hand
does not imply Gaussian dominance since one can imagine
an alternative scenario (to be referred to as
'fine-tuning' picture) where all higher terms in the expansion
(\ref{eq2}) strongly contribute to the lowest $C_D$-proportional
term together with Gaussian correlator while all contributions with
higher powers of $C_D$ and higher Casimirs
effectively cancel each other.
Therefore despite CS property seems to be a very
strong evidence for
Gaussian dominance, the latter stays to be checked
explicitly. See \cite{sdd} in this respect 
where the quartic field correlator
was measured and evidences for its smallness were found.
 We propose here several types of field
correlators which distinguish coherent and
stochastic (i.e. Gaussian) pictures.
The first one is defined as follows:
\be
\lll \T \Phi(x_0,x) F_{\mu\nu}(x) \Phi(x,x_0)
\Phi(x_0,y) F_{\rho\sigma}(y) \Phi(y,x_0) \rrr \equiv
D^{(2)}_{\mu\nu\rho\sigma}
(x - x_0, y-x_0)
\label{cor1}
\ee
The phase factors are defined according to (\ref{transp})
with the straight lines connecting edge points
as integration contours (see Fig.2a).
The $x_0$-dependence
of (\ref{cor1}) is related to the contribution of
correlators higher than Gaussian, but respecting CS.
Gaussian dominance implies smallness of this contribution.
Notice, at the same time
that (\ref{cor1}) is exactly proportional to the quadratic
Casimir by construction.
We take all fields in the expression (\ref{cor1})
in the fundamental representation for the sake of simplicity.
Using standard relations for the phase factors (see, for
example, \cite{nast}) one gets:
$$
\frac{\partial}{\partial x_0^{\alpha}}\> D^{(2)}_{\mu\nu\rho\sigma}
= ig \int\limits_0^1 ds s
(x-x_0)_{\beta} \lll \T F_{\mu\nu}(x) [
F_{\beta\alpha}(x+s(x_0-x)) F_{\rho\sigma}(y)] \rrr
$$
\be
+ ig \int\limits_0^1 ds s
(y-x_0)_{\beta} \lll \T F_{\mu\nu}(x) [
F_{\beta\alpha}(y+s(x_0-y)) F_{\rho\sigma}(y)]
\rrr
\label{cor2}
\ee
with the explicit contributions from the 3-point
correlator to the r.h.s. of (\ref{cor2}) (we omit
$\Phi$'s for brevity).
One can also define two typical lengths, characterizing the
behavior of (\ref{cor1}) in different regions of parameter
space. It is physically clear that the situation when
$|x-y| \sim |x-x_0| + |y-x_0|$ should be distinguished
from the case with the distant reference point
$|x-y| \ll |x-x_0| + |y-x_0|$.
The longitudinal correlation
length $L_{||}$ corresponding to the  variation
of $x$
is to be compared
with the transverse one $L_{\bot}$,
which parametrize the change of (\ref{cor1})
with respect to $x_0$
in both regions in lattice simulations.
It is also possible to choose
rectangular geometry,
it corresponds to the correlator
\be
\lll \T F_{\mu\nu}({\vec 0} , 0) \Phi({\vec 0},0; {\vec z}, 0)
\Phi({\vec z},0; {\vec z}, T) \Phi({\vec z}, T; {\vec 0}, T)
F_{\rho\sigma}({\vec 0}, T)
\Phi({\vec 0},T; {\vec z}, T)
\Phi({\vec z},T; {\vec z}, 0) \Phi({\vec z}, 0; {\vec 0}, 0)
\rrr
\label{qw}
\ee
where all integration contours are again straight lines
(see Fig. 2b).
The $z$--dependence of (\ref{qw}) is to be explored.
The correlators of this form enter different expressions
when one quark is heavy and can be considered as a static source.
In Gaussian dominance scenario one takes into account
$T$-dependence and neglects
$z$--dependence of (\ref{qw})
\cite{sim4}.

The second type of correlators directly distinguishing
stochastic and coherent pictures is given by
multipoint irreducible field strength correlators.
The simplest nontrivial one is that of the fourth order
(see Fig. 2c, where the choice of contours for four-point correlator
with all arguments on straight line is depicted).
One has to consider
the following expression
$$
{\Delta}_D^{(4)}[S] =
\int d\sigma(x_1)
\int\limits^{x_1} d \sigma(x_2)
\int\limits^{x_2} d \sigma(x_3)
\int\limits^{x_3} d \sigma(x_4)
\lll {\T}_{D} F(x_1)
 F(x_2)
 F(x_3)
 F(x_4)\rrr
$$
\be
- \frac12 \> {\left[
\int d\sigma(x_1)
\int\limits^{x_1} d \sigma(x_2)
\lll {\T}_{D} F(x_1)
 F(x_2) \rrr \cdot
\right]}^2
\label{vv2}
\ee
where we omit Lorentz indices
and phase factors $\Phi(x_i, x_j)$.
Notice that in confinement phase where field
correlators develop finite correlation length,
each term in the r.h.s. of (\ref{vv2}) is proportional to $R^2$
at large distances, while their 
difference and hence the l.h.s. of (\ref{vv2})
is linear in $R$.

The ordered integration goes over the surface $S$, parametrized
by $x(s,t): x(0,t)\equiv x_0; x(1,t)\in \partial S$
according to 
$$
\int\limits^{x_1} d \sigma^{\mu\nu}(x_2)\> [...]  \equiv
\int\limits_0^1 ds_2 \int\limits_0^{t_1}
dt_2 \>\left(
\frac{\partial x_2^{\mu}(s_2,t_2)}{\partial s_2}
\frac{\partial x_2^{\nu}(s_2,t_2)}{\partial t_2}
-
\frac{\partial x_2^{\nu}(s_2,t_2)}{\partial s_2}
\frac{\partial x_2^{\mu}(s_2,t_2)}{\partial t_2}
\right)\> [...]
$$
It should be clear
from the previous discussion that in Gaussian scenario
such surface-dependent contribution
${\Delta}_D^{(4)}[S]$ is assumed to be small
in comparison with the two-point one
\be
{\Delta}_D^{(2)}[S] =
\int d\sigma(x_1)
\int\limits^{x_1} d \sigma(x_2)
\lll {\T}_{D} F(x_1)
 F(x_2) \rrr
\label{vv3}
\ee
if the surface $S$ is the minimal
surface.
There exist lattice indications
\cite{sdd} that this is indeed the case.
It would be of considerable
interest to explicitly estimate (\ref{vv2})
in comparison with (\ref{vv3}) on the lattice
using powerful present day lattice technique
and also to study the CS behaviour of
(\ref{vv2}).

The suggested probes being particular examples of
path-dependent gauge-invariant field correlators
are more subtle than Wilson loop averages and
can be effectively used to discriminate
coherent and stochastic pictures in direct way.

\subsection{NP effects at small distances}

It was recently suggested on both
theoretical \cite{zz} and lattice \cite{bali3}
grounds that leading NP contribution
to the static potential at small distances
linearly depended on $R$, instead of the quadratic
dependence naively suggested by conventional OPE:
\be
V_D(R) = -C_D \>\frac{\alpha_s}{R} + {\tilde\sigma}_D \cdot R
\label{ope}
\ee
where the dimension-two coefficient ${\tilde\sigma}_D$
should be distinguished from the asymptotic string
tension $\sigma_D$ in (\ref{pot}). From
the background perturbation theory point of view
described in the previous section, this
linear term comes from the interference between
perturbative and nonperturbative interactions.
In theoretical picture
developed in \cite{zz} the phenomenon was
associated with formation of the so called
infinitely thin string between charges.
One can address the issue of CS
in this approach.
It is of interest to understand theoretically and on the lattice
in this framework the relation between "thick confining strings"
which provide confinement and infinitely "thin short
strings" violating conventional OPE
and presumably
contributing to the linear potential at small distances
in the light of CS in both the small and the large distance regions,
found in \cite{bali2}.

There are arguments to be mentioned
\cite{badalyan} that the quasilinear behavior
of the potential at small distances observed in \cite{bali3}
could be a manifestation of the phenomenon of freezing
of perturbative coupling at large distances.
It is interesting therefore to check
CS in this setup directly, namely to extract not only
${\tilde\sigma}_F$, as it was done in \cite{bali3} but also
the same quantity
for higher $D$ in analogous way.
The linear scaling of ${\tilde\sigma}_D$ with $C_D$
would be in line with
the proposals of \cite{zz} and of course with Gaussian dominance.
Hopefully further lattice investigations
will shed light on that.

\subsection{Interaction of Wilson loops}

Another lattice setup where the CS phenomenon
can be successfully checked is the Wilson loop
correlators. In the present subsection we start from
the disconnected correlators of the
form $\lll \prod\limits_i W_{D_i}(C_i)\rrr $
and come to the
discussion of the connected ones
$ \lll \T_D \prod\limits_i \Phi_D(C_i)\rrr $
after that.
The case of interaction between
loops in different representations is of particular
interest.
We consider the two-loop correlator
in the regime where the sizes of both loops are much smaller
than the distance $R$ between them.
One gets (see Fig.3a)
\be
\lll W_{D_1}(C_1) \cdot W_{D_2}(C_2) \rrr
-
\lll W_{D_1}(C_1) \rrr \cdot \lll W_{D_2}(C_2) \rrr
\sim
[C_{D_1}\cdot C_{D_2} + ...]\cdot \frac{\exp( -
M_{gl} R)}{N_c^2} 
\label{ccc22} \ee
where $M_{gl}$ does not depend on
the choice of representations $D_1, D_2$.
The terms denoted by dots in (\ref{ccc22})
are proportinal to the higher powers of
$C_{D_i}$ and should be suppressed if Gaussian dominance
holds true. Let us mention the results from \cite{trot}
where some indications were presented that
(\ref{ccc22}) holds true (for adjoint--fundamental correlator).
See also \cite{dos} where the
behavior (\ref{ccc22}) was obtained in the
Gaussian stochastic model framework.

Let us consider now gauge-invariant connected correlators.
Lattice simulations demonstrate
the phenomenon of confining string formation.
The important problem of
confinement  dynamics
is to study the string profile.
This question can be addressed for the
string attached to fundamental sources
and to sources in higher representations as well.
One's first instinct is to say that since for
higher representations the string tension is larger
the same should in some sense be true for the
geometrical characteristics of the string, e.g.
its radius. It was already discussed in the
previous section that it is in fact not true.
To see this, let us consider connected
probe of the following form:
\be
\rho^D_{\mu\nu} =
\frac{\lll
\T \Phi_D(C)\Phi P_{\mu\nu}(x) {\Phi}^{\dagger} \rrr}{\lll \T \Phi_D(C)
\rrr}
-1
\label{str1}
\ee
where phase factors $\Phi$, $\Phi^{\dagger}$ connect the
plaquette $P_{\mu\nu}(x)$ with the loop $\Phi_D(C)$
(compare with (\ref{ccc22}).

Since in the continuum limit $\lll P_{\mu\nu}(x)\rrr_{Q\bar Q} \to
a^2 \lll F_{\mu\nu}(x)\rrr_{Q\bar Q}$ where $a$ is lattice link
one gets
\be
\rho^D_{\mu\nu}(x) = a^2 \left[ C_D \cdot \int\limits_{S(C)}
d\sigma_{\alpha\beta}(y)
D_{\alpha\beta\mu\nu}(y,x) + ...\right] + {\cal O}(a^4)
\label{srt2}
\ee
where $S(C)$ is the minimal area surface inside contour
$C$. Omitted terms in the r.h.s. of (\ref{srt2}) denoted by dots
contain higher powers of $C_D$. It is seen at the same time
that the string radius $\left(d\log \rho_{14} / d x\right)^{-1}$
is representation--independent in Gaussian
stochastic model  (and, generally, for arbitrary
CS respecting field ensemble), if the terms proportional
to $C_D$ dominate in (\ref{srt2}). On the contrary
in the absence of CS for (\ref{str1})
the representation dependence
of $\log \rho_{\mu\nu}$ needs not to be factorizable
and radius of the string for such coherent ensemble
would be representation--dependent.

It is therefore
gauge-invariant analysis of the
QCD confining string profile
for charges in higher representations which could
indicate both CS and Gaussian dominance
in a way different from just static potential
measurements. Notice, that in
\cite{trot,dos} another (disconnected)
type of correlator was used and
the results for (\ref{srt2}) should not be directly
compared to that of those papers.

\subsection{CS for nonstatic
potentials and hadron spectra}

Till the present moment we have considered only static
potentials. From hadron phenomenology point of view,
however, not static but dynamical potentials, which take
into account the effects of quark motion are of interest.
What is the meaning of CS in this case?
 Let us consider for simplicity gauge-invariant
Green's function for heavy-light spinless
system in the gluon field (see, e.g. \cite{lisb})
$$
G(x,y) = {\cal N}\>\int\limits_0^{\infty} ds \> \int
{\cal D} z \> \exp\left[
-m^2 s - \int\limits_0^{s} d\sigma \frac{z_{\mu}^2(\sigma)}{4}
\right] \cdot
$$
\be \cdot
\left\lll {\T}_D
{\mbox{P}}\exp \left[
+ i \int\limits_x^y A_{\mu}^a(\vec z, z_4) T^a dz_{\mu}
\right]\cdot
{\mbox{P}}\exp \left[
- i \int\limits_{x_4}^{y_4} A_{4}^a(\vec 0, t) T^a dt
\right]
\right\rrr
\label{fs1}
\ee
There are several different regimes of (\ref{fs1})
depending on relative value of the light particle
mass $m$ and properties of the gluon ensemble, encoded
in the weight used in the averaging procedure $\lll .. \rrr$.
In case of large $m$ when the nonrelativistic expansion
is valid, one has to distinguish two situations:
the potential regime and the sum-rule limit.
The energy levels get a correction in the latter case
which in Gaussian stochastic model is proportional to $C_D$.
Contributions of higher condensates are suppressed
in this case by the inverse powers of $m$.
In the potential regime the situation becomes less transparent
and one is not to expect any exact scaling law with respect to
$C_D$ for the
spectrum of the system, despite the potential obeys CS. 
Spin degrees of freedom provide additional
complications. In particular, difference in spin
structures precludes to identify the spectrum of the
heavy-light system where spinor light particle carries
fundamental color charge (which
models $D$ and $B$-mesons) with heavy-light spectrum
where vector-like dynamical constituent carries
adjoint colour charge (such system might mimic
hypothetical hybrid excitations of heavy mesons).
Indeed, the only direct fingerprint of CS for the
real particles spectrum is presumably the fact that
the lowest glueballs are believed to be heavier than
the lowest mesons (we do not take chiral effects
into account). 
Let us consider "hadrons" as bound states of massless and spinless
quarks, without perturbative interactions.
In this case the mass of a hadron is proportional
to the square root of $\sigma_D$ - the only dimensionful
parameter of the problem: $M_n = c_n \cdot \sqrt{\sigma_D} \sim
\sqrt{C_D}$. Hence one might predict that the equivalent
states in glueball and meson spectrum have ratio of masses
$M^{gg} / M^{q\bar q} = \sqrt{9/4} = 3/2$.
Now include spins and perturbative dynamics.
Taking into account of Coulomb force changes this ratio,
also spin-dependent forces are different since spin of gluon
is twice that of quark, and finally quarks have (negative)
self-energy corrections (of the order -0.25 GeV per particle
\cite{simnn})
while for gluons selfenergies are forbidden by gauge invariance.
All that leads to the masses calculated in potential model
in \cite{kaid} which are in good agreement with lattice and experimental
data, namely for center-of-mass of $2^{++}$ and $0^{++}$ glueball
one finds $M^{gg(L=0)}_{com} \equiv \frac12
(M^{gg} (2^{++}) + M^{gg} (0^{++})) \approx 2$ Gev, while
$m_{\rho} = 0.77$ GeV and similarly for $M^{gg(L=1)}_{com}= 2.66$
GeV, while for mesons $M^{qq(L=1)}_{com} = 1.2$ GeV.
Ratio of glueball to meson masses in both cases is around
$2.2 - 2.5$, which is larger than 3/2 = 1.5. The main reason is
quark selfenergies and different role of Coulomb
forces: while for $\bar q q $ those are significant, for
$gg$ system as argued in \cite{kaid} perturbative gluon exchanges
are suppressed due to higher loop effects.

\subsection{CS and QCD sum rules.}

The OPE series which is the starting point of QCD
sum rules contains two types of higher twist term: those which
contain color-irreducible combinations of field operators,
like $\lll D^k F_{\mu\nu} D^n F_{\sigma\lambda}\rrr $ or
$\lll \bar\psi \sigma_{\mu\nu} F_{\mu\nu} \psi \rrr$
and reducible operators like $\lll \bar\psi \psi \bar\psi \psi \rrr$
or $\lll FFFF\rrr$. For the latter the conventional strategy
\cite{svz} is the so called vacuum insertions, replacing
e.g. $\lll FFFF\rrr \to \lll FF\rrr \lll FF\rrr$.
This replacement procedure was however criticized
from the point of view of the instanton gas/liquid
model, which does not suport good accuracy of vacuum insertions.
Now after measurements done in \cite{bali2,deldar1}
and subsequent analysis in \cite{we1,we2} one can say
more about accuracy of vacuum insertions at least in
purely gluonic operators. Namely, such vacuum insertion
is equivalent to neglection of higher irreducible correlators
(cumulants), e.g. $\llll FFFF \rrrr \sim
\lll FFFF\rrr - \lll FF\rrr \lll FF\rrr \approx 0$
and since quartic and higher order correlators violate
CS, their admixture is limited by the measured accuracy of CS,
i.e. of a few percents. The same CS also shows (see
Section III of this paper) that instanton model of vacuum
fails to reproduce the correct behavior of static potential
and hence the procedure of vacuum insertions for gluonic operators
is fully justified with shown above accuracy. Roughly
speaking, CS favours vacuum insertions.

Another facet of that is also worth mentioning.
From the spectral decomposition point of view
the vacuum insertion procedure is based on the fact that
there is a mass gap in the theory and a set of
massive (colorless) states gives small contribution to
the spectral decomposition of color-reducible correlator.
On the other hand, vacuum is made of
colorless weakly interacting dipoles in Gaussian picture
and confinement (and hence the mass gap) is associated
with the fact that these dipoles have finite (and actally small)
size. The size of these dipoles is inversly proportional to
the lowest gluelump mass (see above).
On the contrary,
the formal OPE limit is achieved by taking this size
as infinite, $T_g \to \infty$ in terms of FCM,
when field correlators become
just $x$-independent condensates. It is of interest to clarify this problem
selfconsistently and to establish the form of the operator product expansion in
the physical Gaussian vacuum, characterized by the finite correlation length.
This work is in progress now.

\subsection{CS at $T\neq 0$ and $\mu \neq 0$ }

Lattice simulations of QCD at nonzero temperature
could provide additional information about the discussed question.
One should distinguish the Polyakov loop, i.e. trace
of the phase factor taken along temporal axis from
$0$ to $\beta = 1/T$, and usual Wilson loop, which is made of
 spacial links. It is known (see e.g. \cite{ds3}
and references therein) that at the point of deconfinement
phase transition the magnetic components of the
chromofields stay approximately
intact while the electric components responsible
for the nonzero string tension of the temporal Wilson loop vanish.
In accordance to that we expect that the CS law for the potential
extracted from the spacial loops not to be spoiled
at $T>0$ and even at $T>T_c$. As for the Polyakov loops,
the CS has no simple meaning for the averaged
traces of temporal links. At $T<T_c$ the
Polyakov loops for nonzero triality charges vanish
while the screening happens for other
representations. Therefore there is no room for CS
region in this case. After the deconfinement transition
one comes back to the dynamical CS and has to study
the spectrum of heavy-light lumps with the heavy source
taken to belong to the given representation.
As it was argued in the previous subsection,
such spectra need not necessarily obey CS.
At the same time it is important to notice that
lattice shows \cite{gross} one and the same deconfinement
temperature extracted from Polyakov loops in different
representations. This fact seems to have no direct relation with
CS, at the same time, it is very natural in Gaussian dominance
picture -- it just indicates that the
electric part of the two-point field strength correlator,
responsible for nonzero string tension, vanishes
above the critical temperature. The actual value of the
deconfinement temperature does not depend
on the properties of external current.
It would however be very interesting
if lattice simulations will indicate that
CS is temperature-dependent, studying
either spacial or temporal loops.
In Gaussian dominance picture it will signalize
that the dynamics governed by higher correlators
is different at different temperatures.

Another closely related problem is the
determination of
higher representation Polyakov
loop behaviour when $T\to T_c $ from above
(see, e.g. \cite{kiskis}, \cite{damg}).
The difference in critical exponents
for higher representation Polyakov
loops near the point of the phase transition
presumably can be interpreted
in terms of some bounds on higher cumulant
contributions in (\ref{eq2}).
This set of questions in connection with CS
will be discussed elsewhere.

The situation at $\mu \neq 0$ is much more subtle.
It is supposed \cite{colo1} that at high density
the phenomenon of color superconductivity takes place.
One should distinguish 
two and three flavor cases. In the latter case the
so called color-flavor locking leads 
(see review \cite{colo2} and references therein)
to nonzero
and equal masses for all eight adjoint 
colors of gluons, while in the two-flavor case 
vacuum state breaks global color symmetry $SU(3) \to SU(2)$
in such a way that three gluons corresponding 
to unbroken gauge subgroup $SU(2)$ remain massless while 
five others take nonzero and unequal 
(in low temperature phase)
masses.
The picture is therefore in some correspondence with the
Georgi-Glashow
model discussed above and it is reasonable to expect 
that CS is 
destroyed in the region of densities corresponding to two-flavor
color 
superconductivity (and restored back again 
at densities 
where the third flavor comes into play). 
Despite external Wilson loop seems to be rather 
artificial probe at high density and 
lattice simulations
of $N_c = 3$ QCD at $\mu \neq 0$ are unfortunately 
impossible at the moment, it is tempting to 
speculate that two-flavor color superconductor if exists
might occupy the only region on the QCD phase diagram where 
the CS property is broken in explicit way.

\section{Conclusion and outlook}

Accurate lattice data \cite{bali1,bali2,deldar1} on the static
potential for the sources in higher $SU(3)$
representations strongly support the Casimir
scaling (CS) hypothesis \cite{ambjorn}. Most of the
popular models of confinement encounter
internal difficulties in attempts to explain this fact.
We have shown in the present article
that CS strongly supports the Gaussian dominance
of QCD vacuum.
Let us summarize the paths which we propose
for further studies of this complicated dynamical problem.

\begin{itemize}
\item To study CS of perturbative series beyond two loops
both in continuum limit and on the lattice.
\item To investigate CS in the course of the
cooling procedure.
\item To establish exact results for the static
potentials between charges in higher representations
in SUSY QCD and compare them with CS-respecting
results of ordinary QCD.
\item Convincing evidences for the string breaking phenomenon
will allow to establish quantitatively the bounds of
the CS region.
\item CS might break at large enough distances
were the quantum dynamics of the confining string starts
to be detectable. Needless to say that it
would be very interesting to observe the regime
of vibrating confining string, if it exists.
\item The set of lattice measurements which could significantly
improve our understanding of CS and Gaussian dominance
includes
static 3--point correlators,
nonperturbative potential at small distances,
string profile, Wilson loop correlators in higher
representations etc.
\item It should be stressed again that despite
      Gaussian dominance straightforwardly implies CS,
the opposite strictly speaking is not true.
By direct measurements of path-dependent
field strength correlators one can check explicitly
the accuracy of Gaussian dominance (i.e. comparing 
(\ref{vv2}) and (\ref{vv3})).

\end{itemize}

Whatever the dynamical reason of CS is,
its persistence strongly advocates that QCD vacuum
is stochastic rather than coherent. It provides also
some limits on the use of instanton gas/liquid model and other
models of confinement based on classical solutions.
As was explained above stochastic picture of QCD vacuum
incorporates some features of the dual Meissner effect.

In terms of field correlator method
Gaussian dominance implies the dominant NP
contribution to the gauge-invariant
observables from the NP parts of the two-point field
strength correlator. Physically it corresponds
to the picture of the vacuum, made of relatively
small white dipoles. The size of these dipoles
which is natural to associate with $T_g$
controls the stochasticity of the vacuum, i.e. the
spacial size of the domains where fields are coherent
in the spirit of old ideas \cite{olesen}. We need however
more than that -- indeed, Gaussian dominance implies
the approximate ideality of this "dipole gas", i.e.
relatively weak interaction between dipoles and
as a consequence a small fraction of
the higher "multipoles" in the vacuum.
One can possibly understand this structure solving
nonlinear equations for field correlators generalized
by inclusion of perturbative and higher correlator terms.

The discussed white "dipoles" live on the surfaces bound
by Wilson contours and the relative integral weight
of higher irreducible cumulants ("multipoles")
depends on the shape of these surfaces
despite total answer is of course
surface-independent and the dynamics of the dipole
ensemble is topological in this sense.
Gaussian dominance means that one can find such
surface $S$ -- the minimal one --
where the dipoles
effectively become quasifree.
Notice the difference from the
dimensional reduction scenario
where the effective $2d$ surface is populated
by colored vector particles --
gluons -- with no any interaction between
them for arbitrary geometry of the surface.

Finally, the greatest mystery which probably explains
all other features is the small value of gluon
correlation length $T_g$ (or, equivalently,
large value of the lowest gluelump
mass, which is around 1.5 Gev) in units of
string tension. Taken together with even
higher values for multigluon gluelumps, yielding the
correlation lengths of higher cumulants, this might explain
both the Gaussian dominance and the high accuracy of CS.

\acknowledgements
The authors are grateful to G.S. Bali,
S.Deldar, A.Yu.Dubin, M.Ilgenfritz and
L.B.Okun for valuable discussions.
One of the
authors (Yu.S.) acknowledges the kind hospitality
of the Institute for Theoretical Physics of
Utrecht University and
useful discussions with N. Van Kampen,
G.'t Hooft and J.A.Tjon.

This work was partially supported by the grants
RFFI 00-02-17836 and
RFFI 00-15-96786 for scientific
schools. V.Sh. acknowledges the support from the
grant RFFI 01-02-06284.

\newpage

\begin{center}

{\bf \large Figures}

\bigskip

\begin{picture}(400,220)
\put(110,110){\line(0,1){60}}
\put(110,110){\line(2,-1){53}}
\put(110,110){\line(-2,-1){53}}
\put(115,175){$Q$}
\put(50,72){$Q$}
\put(165,72){$Q$}
\put(290,170){\line(3,-5){53}}
\put(290,170){\line(-3,-5){53}}
\put(237,82){\line(1,0){106}}
\put(295,175){$G$}
\put(222,75){$G$}
\put(349,75){$G$}
\put(100,40){{\bf Fig.1a}}
\put(280,40){{\bf Fig.1b}}
\end{picture}

\begin{picture}(500,200)
\put(80,70){\line(0,1){60}}
\put(80,130){\line(1,0){15}}
\put(95,130){\line(0,-1){15}}
\put(95,115){\line(-1,0){12}}
\put(83,115){\line(0,-1){48}}
\put(83,67){\line(-1,0){66}}
\put(17,67){\line(0,1){15}}
\put(17,82){\line(1,0){15}}
\put(32,82){\line(0,-1){12}}
\put(32,70){\line(1,0){48}}
\put(87,60){$x_0$}
\put(17,90){$F(x)$}
\put(80,139){$F(y)$}
\put(180,70){\line(0,1){60}}

\put(220,130){\line(1,0){15}}
\put(180,130){\line(-1,0){15}}
\put(235,130){\line(0,-1){15}}
\put(165,130){\line(0,-1){15}}
\put(235,115){\line(-1,0){12}}
\put(165,115){\line(1,0){12}}

\put(223,115){\line(0,-1){48}}
\put(177,115){\line(0,-1){48}}
\put(177,67){\line(1,0){46}}

\put(220,70){\line(0,1){60}}
\put(180,70){\line(1,0){40}}
\put(160,57){$({\vec z}, 0)$}
\put(225,57){$({\vec z}, T)$}
\put(160,140){$F({\vec 0}, 0)$}
\put(215,140){$F({\vec 0}, T)$}
\put(190,30){{\bf Fig.2b}}
\put(45,30){{\bf Fig.2a}}
\put(358,30){{\bf Fig.2c}}
\put(300,60){\line(1,0){135}}
\put(300,60){\line(0,1){15}}
\put(435,60){\line(0,1){15}}
\put(300,75){\line(1,0){15}}
\put(435,75){\line(-1,0){15}}
\put(315,75){\line(0,-1){12}}
\put(315,63){\line(1,0){25}}
\put(340,63){\line(0,1){12}}
\put(340,75){\line(1,0){15}}
\put(355,75){\line(0,-1){12}}
\put(355,63){\line(1,0){25}}
\put(380,63){\line(0,1){12}}
\put(380,75){\line(1,0){15}}
\put(395,75){\line(0,-1){12}}
\put(395,63){\line(1,0){25}}
\put(420,63){\line(0,1){12}}
\put(295,82){$F(x_1)$}
\put(335,82){$F(x_2)$}
\put(375,82){$F(x_3)$}
\put(415,82){$F(x_4)$}
\end{picture}

\begin{picture}(400,160)
\put(20,20){\line(1,0){60}}
\put(80,20){\line(0,1){60}}
\put(20,20){\line(0,1){60}}
\put(20,80){\line(1,0){60}}
\put(100,30){\line(1,0){40}}
\put(140,30){\line(0,1){40}}
\put(100,30){\line(0,1){40}}
\put(100,70){\line(1,0){40}}
\put(45,90){$W_1$}
\put(115,80){$W_2$}
\put(220,20){\line(1,0){83}}
\put(280,80){\line(0,-1){57}}
\put(220,80){\line(0,-1){60}}
\put(220,80){\line(1,0){60}}
\put(245,90){$W_D$}
\put(308,65){$P_{\mu\nu}$}
\put(303,30){\line(1,0){22}}
\put(325,30){\line(0,1){25}}
\put(300,55){\line(0,-1){32}}
\put(300,55){\line(1,0){25}}
\put(280,23){\line(1,0){20}}
\put(303,30){\line(0,-1){10}}
\put(75,0){{\bf Fig.3a}}
\put(262,0){{\bf Fig.3b}}
\end{picture}
\end{center}

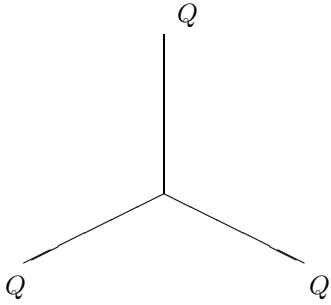
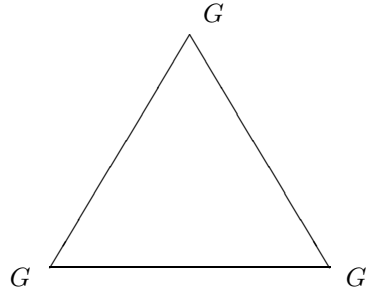
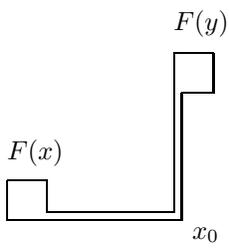
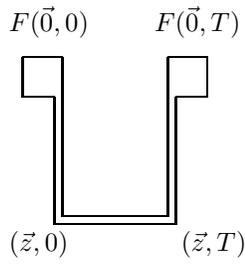
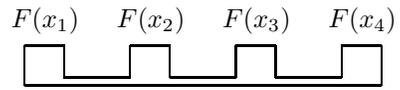
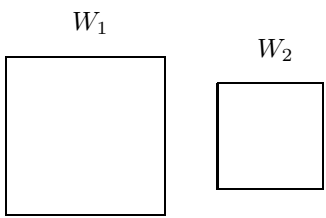
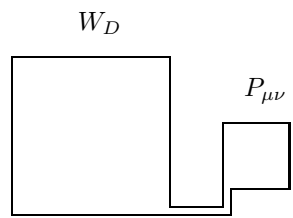
\end{document}